\documentclass[eqsecnum,twocolumn,aps,epsf]{revtex4}
\usepackage{graphicx}
\usepackage{color}
\begin{document}
\draft

\title{Effects of Zn and Ni substitution on the Cu-spin dynamics and superconductivity in La$_{2-x}$Sr$_x$Cu$_{1-y}$(Zn,Ni)$_y$O$_4$ with $x=0.15 - 0.20$ studied by the muon spin relaxation and magnetic susceptibility}

\author{T. Adachi}
\thanks{Corresponding author: adachi@teion.apph.tohoku.ac.jp}
\author{N. Oki}
\author{Risdiana}
\altaffiliation[Present address: ]{Department of Physics, Faculty of Mathematics and Natural Sciences, Padjadjaran University, Jl. Raya Bandung-Sumedang Km. 21 Jatinangor, West Jawa, Indonesia 45363.}
\author{S. Yairi}
\author{Y. Koike}
\affiliation{Department of Applied Physics, Graduate School of Engineering, Tohoku University, 6-6-05 Aoba, Aramaki, Aoba-ku, Sendai 980-8579, Japan}

\author{I. Watanabe}
\affiliation{Advanced Meson Science Laboratory, RIKEN Nishina Center, 2-1 Hirosawa, Wako 351-0198, Japan}

\date{\today}

\begin{abstract}
We have investigated effects of Zn and Ni on the Cu-spin dynamics and superconductivity from the zero-field muon-spin-relaxation (ZF-$\mu$SR) and magnetic-susceptibility, $\chi$, measurements for La$_{2-x}$Sr$_x$Cu$_{1-y}$(Zn,Ni)$_y$O$_4$ with $x = 0.15 - 0.20$, changing $y$ up to 0.10 in fine step. 
In the optimally doped $x=0.15$, it has been concluded that the formation of a magnetic order requires a larger amount of Ni than that of Zn, which is similar to our previous results of $x = 0.13$. 
From the estimation of volume fractions of superconducting (SC) and magnetic regions, it has been found for $x=0.15$ that the SC region is in rough correspondence to the region where Cu spins fluctuate fast beyond the $\mu$SR frequency window for both Zn- and Ni-substituted samples. 
According to the stripe model, it follows that, even for $x = 0.15$, the dynamical stripe correlations of spins and holes are pinned and localized around Zn and Ni, leading to the formation of the static stripe order and the suppression of superconductivity. 
These may indicate an importance of the dynamical stripe in the appearance of the high-$T_{\rm c}$ superconductivity in the hole-doped cuprates. 
In the overdoped regime of $x=0.18$ and 0.20, on the other hand, the SC region seems to be in rough correspondence to the region where Cu spins fluctuate fast beyond the $\mu$SR frequency window, though it appears that the Cu-spin dynamics and superconductivity are affected by the phase separation into SC and normal-state regions. 

\end{abstract}
\vspace*{2em}
\maketitle
\newpage

%*****************************************************************************************
\section{Introduction}\label{intro}
%*****************************************************************************************
Toward understanding of the mechanism of high-$T_{\rm c}$ superconductivity, an approach through the substitution of impurities for Cu has been one of major ways to study the Cu-spin dynamics in the high-$T_{\rm c}$ physics. 
In the hole-doped high-$T_{\rm c}$ cuprates, it is well known that nonmagnetic impurities such as Zn tend to suppress the superconductivity more markedly than magnetic impurities such as Ni,~\cite{xiao} which is an opposite trend to the conventional superconductors.~\cite{tinkham} 
Early NMR and NQR measurements have revealed that Zn acts as a strong scatterer of holes causing the unitarity scattering, while Ni acts as a weak scatterer causing the Born scattering.~\cite{kitaoka} 
From inelastic neutron-scattering experiments focusing on the high-energy Cu-spin fluctuations, it has been reported that the so-called resonance energy does not change through the Zn substitution, while it decreases through the Ni substitution.~\cite{sidis} 
As for the low-energy Cu-spin fluctuations, on the other hand, Kimura {\it et al}.~\cite{kimura} and Kofu {\it et al}.~\cite{kofu} have insisted from the inelastic neutron-scattering measurements in La$_{2-x}$Sr$_x$Cu$_{1-y}$(Zn,Ni)$_y$O$_4$ with $x = 0.15$ that Zn tends to bring about the so-called in-gap state in the spin gap, while Ni tends to reduce the spin gap itself. 
These results are strongly suggestive of different effects between the Zn and Ni substitution on the Cu-spin dynamics and superconductivity, but the reason has not yet been clarified.

Formerly, we have performed zero-field (ZF) muon-spin-relaxation ($\mu$SR) and magnetic-susceptibility, $\chi$, measurements in La$_{2-x}$Sr$_x$Cu$_{1-y}$(Zn,Ni)$_y$O$_4$ around $x = 0.115$ at low temperatures down to 2 K, changing $y$ up to 0.10 in fine step,~\cite{nabe-prb,nabe-jpcs,nabe-physb,ada-jltp,ada-prbz,ada-prbn} in order to investigate impurity effects on the so-called spin-charge stripe order~\cite{tranquada} which may be related to the mechanism of high-$T_{\rm c}$ superconductivity.~\cite{kivelson} 
For the Zn substitution, a muon-spin precession corresponding to the formation of a long-range magnetic order has been observed for $y$(Zn) $=0.0075 - 0.03$ in $x=0.10$, 0.115, 0.13.~\cite{ada-prbz} 
Moreover, the further Zn substitution of $y$(Zn) $>0.03$ brings about a change to slow depolarization of muon spins so that Gaussian-like depolarization has been observed for $y$(Zn) $=0.10$.
These results indicate that a magnetic order is stabilized by a slight amount of Zn, while it is destroyed by a large amount of Zn, which is understandable in terms of the pinning and destruction of the stripe order by Zn.~\cite{koike-pin,koike-pin2,ada-pin}
For the Ni substitution, on the other hand, a muon-spin precession has been observed for $y$(Ni) $\ge 0.02$ up to 0.10 in $x=0.13$.~\cite{ada-prbn} 
These indicate that the formation of a magnetic order requires a larger amount of Ni than that of Zn and that the magnetic order survives, or rather, it is stabilized by a large amount of Ni. 
From the $\chi$ measurements on field cooling, it has been found that the Meissner volume fraction corresponding to the superconducting (SC) volume fraction rapidly decreases with a slight substitution of Zn and its decrease is more marked by Zn than by Ni.
In comparison between the volume fraction of the SC state estimated from $\chi$ and that of the magnetic state from $\mu$SR, it has been found that the SC region in a sample is in good correspondence to the region in which Cu spins fluctuate fast beyond the $\mu$SR frequency window ($10^6 - 10^{11}$ Hz).
From the viewpoint of impurity effects on the stripe order, these results suggest that (i) the dynamical stripe correlations tend to be pinned and stabilized by both Zn and Ni so that the superconductivity is destroyed around themselves, and that (ii) Zn is more effective for the pinning of stripes so that the static stripe-ordered region around Zn is larger than around Ni.
These may be the reason why Zn destroys the superconductivity in the hole-doped high-$T_{\rm c}$ cuprates more markedly than Ni.~\cite{xiao} 

In this paper, in order to investigate the relationship between the Cu-spin dynamics and superconductivity in the optimally doped high-$T_{\rm c}$ cuprates, we have performed ZF-$\mu$SR and $\chi$ measurements in the Zn- or Ni-substituted La$_{2-x}$Sr$_x$Cu$_{1-y}$(Zn,Ni)$_y$O$_4$ with $x = 0.15$, changing $y$ up to 0.10 in fine step.~\cite{koike,ada-m2s} 
Moreover, we have extended the measurements for the Zn-substituted La$_{2-x}$Sr$_x$Cu$_{1-y}$Zn$_y$O$_4$ up to $x=0.20$, focusing on the relationship between the Cu-spin dynamics and superconductivity in the overdoped regime. 

%*****************************************************************************************
\section{Experimental}
%******************	***********************************************************************
Polycrystalline samples of La$_{2-x}$Sr$_x$Cu$_{1-y}$(Zn,Ni)$_y$O$_4$ with $x = 0.15 - 0.20$ and $y=0 - 0.10$ were prepared by the ordinal solid-state reaction method.
Details of the procedure have been reported in our former paper.~\cite{ada-prbz}
For the Ni-substituted samples, however, the sintering was done in air at 1150 $^{\rm o}$C for 24 h to keep the homogeneity of constituents in a sample.

The ZF-$\mu$SR measurements were performed at the RIKEN-RAL Muon Facility at the Rutherford-Appleton Laboratory in the UK, using a pulsed positive surface muon beam with an incident muon momentum of 27 MeV/c. 
The asymmetry parameter $A(t)$ at a time $t$ was given by $A(t)=\{F(t)-\alpha B(t)\}/\{F(t)+\alpha B(t)\}$, where $F(t)$ and $B(t)$ are total muon events of the forward and backward counters, which were aligned in the beam line, respectively. 
The $\alpha$ is the calibration factor reflecting the relative counting efficiencies between the forward and backward counters. 
The $\mu$SR time spectrum, namely, the time evolution of $A(t)$ was measured at low temperatures down to 0.3 K to detect the appearance of a magnetic order. 

The $\chi$ measurements were carried out down to 2 K using a SQUID magnetometer (Quantum Design, Model MPMS-XL5) in a magnetic field of 10 Oe on field cooling, in order to evaluate the volume fraction of the SC region in a sample. 

%*****************************************************************************************
\section{Results}
%*****************************************************************************************
\subsection{Effects of Zn and Ni substitution in $x=0.15$}
%*****************************************************************************************
\subsubsection{ZF-$\mu$SR}
%*****************************************************************************************
\begin{figure*}[tbp]
\begin{center}
\includegraphics[width=0.7\linewidth]{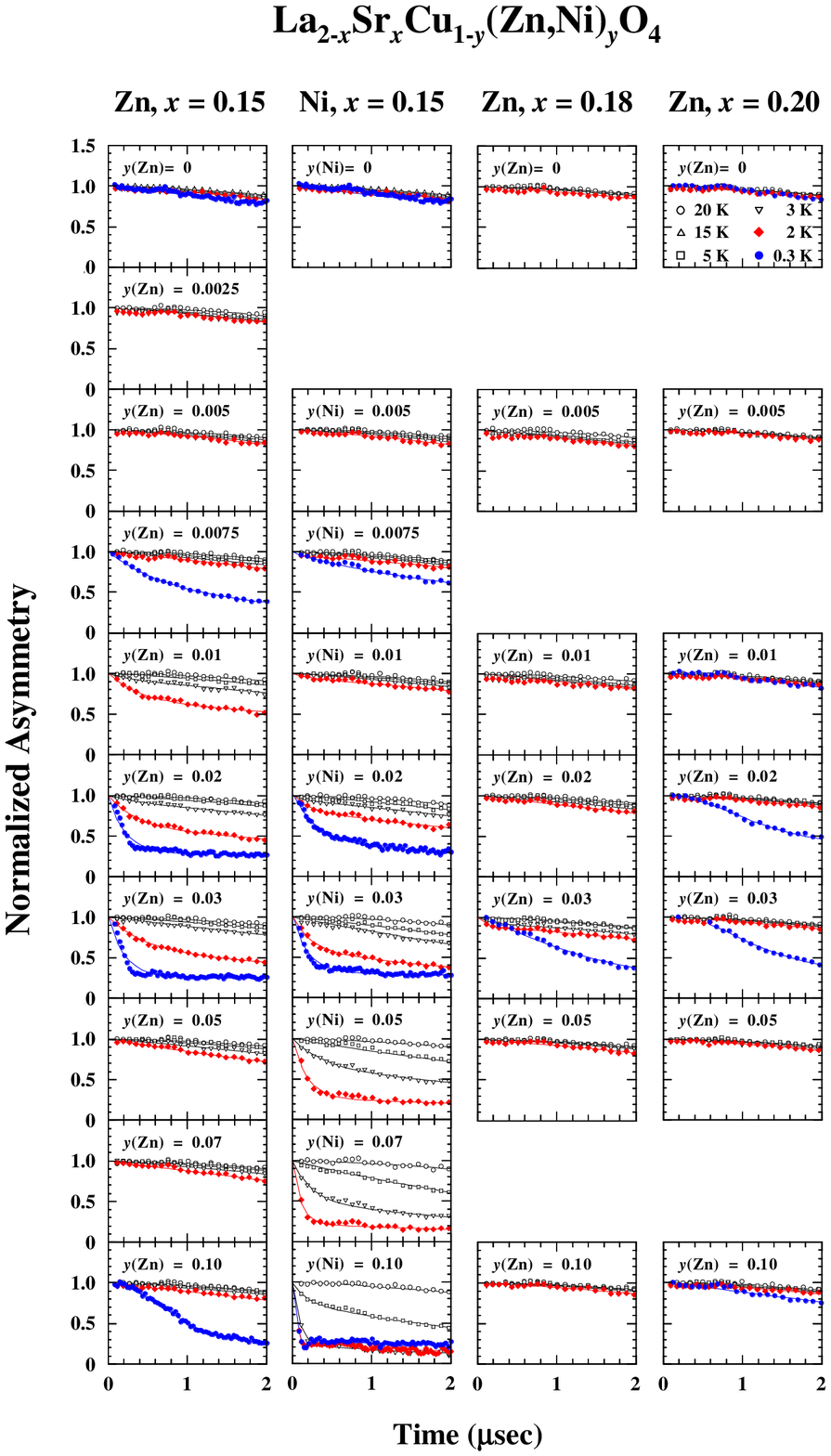}
\end{center}
\caption{(color online) Zero-field $\mu$SR time spectra in the early time region from 0 to 2 $\mu$s of La$_{2-x}$Sr$_x$Cu$_{1-y}$(Zn,Ni)$_y$O$_4$ with $x=0.15$, 0.18 and 0.20 at various temperatures down to 0.3 K. Solid lines indicate the best-fit results using $A(t) = A_0 e^{-\lambda_0t}G_Z(\Delta,t) + A_1 e^{-\lambda_1t} + A_2 e^{-\lambda_2t}{\rm cos}(\omega t + \phi)$ except for $y$(Zn) $=0.10$ in $x=0.15$. For $y$(Zn) $=0.10$ in $x=0.15$, the following equation is used, $A(t) = A_0 e^{-\lambda_0t}G_Z(\Delta,t) + A_{\sigma} e^{-(\sigma t)^2}$.}  
\label{fig:fig1} 
\end{figure*}

\begin{figure}[tbp]
\begin{center}
\includegraphics[width=1.0\linewidth]{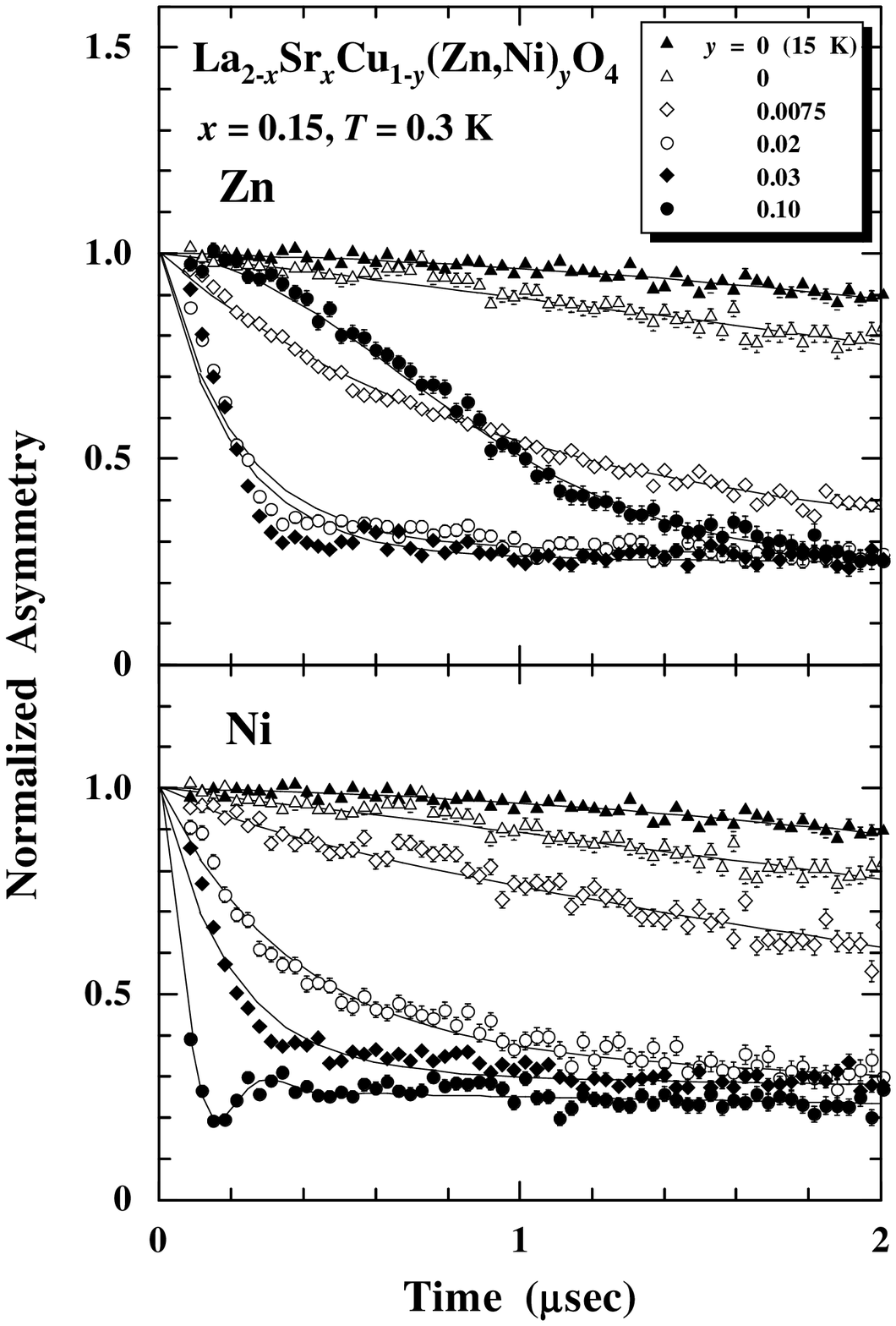}
\end{center}
\caption{Zero-field $\mu$SR time spectra at 0.3 K for typical $y$ values of La$_{2-x}$Sr$_x$Cu$_{1-y}$(Zn,Ni)$_y$O$_4$ with $x=0.15$. The time spectrum at 15 K for $y=0$ is also plotted. Solid lines indicate the best-fit results using $A(t) = A_0 e^{-\lambda_0t}G_Z(\Delta,t) + A_1 e^{-\lambda_1t} + A_2 e^{-\lambda_2t}{\rm cos}(\omega t + \phi)$ except for $y$(Zn) $=0.10$. For $y$(Zn) $=0.10$, the equation, $A(t) = A_0 e^{-\lambda_0t}G_Z(\Delta,t) + A_{\sigma} e^{-(\sigma t)^2}$, is used.} 
\label{fig:fig2} 
\end{figure}

Figure 1 shows the ZF-$\mu$SR time spectra in La$_{2-x}$Sr$_x$Cu$_{1-y}$(Zn,Ni)$_y$O$_4$ with $x=0.15$. 
In order to see the $y$ dependence of the spectra at the lowest temperature of 0.3 K clearly, the time spectra at 0.3 K are also shown in Fig. 2. 
All the spectra in Figs. 1 and 2 is shown after subtracting the background from the raw spectra and being normalized by the value of the asymmetry at $t=0$. 
At high temperatures above 15 K, all the spectra show Gaussian-like slow depolarization of muon spins due to the nuclear-dipole field randomly distributed at the muon site. 
This indicates no effect of Cu spins on the $\mu$SR time spectrum. 
Focusing on the spectra below 2 K, Gaussian-like depolarization is still observed down to 0.3 K for $y=0$, indicating that Cu spins fluctuate fast beyond the $\mu$SR frequency window. 
For the Zn substitution, the muon-spin depolarization becomes fast with increasing $y$(Zn) for $y$(Zn) $\ge 0.0075$. 
Moreover, an almost flat spectrum with the normalized asymmetry of $\sim 1/3$ is observed above 0.4 $\mu$s for $y$(Zn) $=0.02$ and 0.03, indicating the formation of a static magnetic state. 
It is noted that the absence of a muon-spin precession may be due to a short-range magnetic order, resulting in very fast damping of the precession, in contrast to the results around $x = 0.115$.~\cite{nabe-prb,nabe-jpcs,nabe-physb,ada-jltp,ada-prbz,ada-prbn} 
For $y$(Zn) $ > 0.03$, the muon-spin depolarization becomes slow and a Gaussian-like spectrum is observed, though the muon-spin depolarization is still fast at 0.3 K for $y$(Zn) $=0.10$. 
This indicates that the magnetic order is destroyed by a large amount of Zn, which is analogous to the results around $x = 0.115$.~\cite{nabe-prb,nabe-jpcs,nabe-physb,ada-jltp,ada-prbz,ada-prbn}
For the Ni substitution, on the other hand, fast depolarization is observed for $y$(Ni) $\ge 0.0075$. 
Eventually, a muon-spin precession is observed for $y$(Ni) $=0.10$, indicating the formation of a long-range magnetic order of Cu spins. 

As for the comparison of the spectra between the Zn and Ni substitution, the fast depolarization of muon spins starts to be observed at $y = 0.0075$ both in the Zn- and Ni-substituted samples. 
As shown in Fig. 2, however, the muon-spin depolarization at 0.3 K is faster in the 0.75 \% Zn-substituted sample than in the 0.75 \% Ni-substituted one. 
These results indicate that the formation of the magnetic order requires a larger amount of Ni than that of Zn, which is similar to our previous results of $x = 0.13$.~\cite{ada-prbn} 
In the heavily impurity-substituted samples, Zn tends to destroy the Cu-spin correlation, while Ni tends to stabilize the magnetic order, which is also similar to our previous results of $x = 0.13$.~\cite{ada-prbn} 

In order to obtain detailed information on the Cu-spin dynamics, we analyzed the time spectra shown in Figs. 1 and 2 using the following three-component function.
\begin{equation}
A(t) = A_0 e^{-\lambda_0t}G_Z(\Delta,t) + A_1 e^{-\lambda_1t} + A_2 e^{-\lambda_2t}{\rm cos}(\omega t + \phi).
\label{eq1}
\end{equation}
The first term represents the slowly depolarizing component in a region where the Cu spins fluctuate fast beyond the $\mu$SR frequency window. 
The $A_0$ and $\lambda_0$ are the initial asymmetry and depolarization rate of the slowly depolarizing component, respectively. 
The $G_{\rm Z}(\Delta,t)$ is the static Kubo-Toyabe function with a half width, $\Delta$, describing the distribution of the nuclear-dipole field at the muon site.~\cite{uemura} 
The second term represents the fast depolarizing component in a region where the Cu-spin fluctuations slow down and/or a short-range magnetic order of Cu spins is formed. 
The $A_1$ and $\lambda_1$ are the initial asymmetry and depolarization rate of the fast depolarizing component, respectively.
The third term represents the muon-spin precession in a region where a long-range magnetic order of Cu spins is formed. 
The $A_2$ is the initial asymmetry. 
The $\lambda_2$, $\omega$ and $\phi$ are the damping rate, frequency and phase of the muon-spin precession, respectively. 
Values of $A_0$, $A_1$ and $A_2$ obtained from the analysis using Eq. (\ref{eq1}) are regarded as corresponding to volume fractions of each region depending on the frequency of the Cu-spin fluctuations. 

As for $y$(Zn) $=0.10$, the spectra below 1 K cannot be well represented using Eq. (\ref{eq1}) because of the rather fast Gaussian-like depolarization compared with that due to the nuclear-dipole field. 
Therefore, the following function was used.
\begin{equation}
A(t) = A_0 e^{-\lambda_0t}G_Z(\Delta,t) + A_{\sigma} e^{-(\sigma t)^2}.
\label{eq2}
\end{equation}
The first term is the same as that in Eq. (\ref{eq1}).
The second term is a simple Gaussian representing a region with larger magnetic moments than nuclear moments being randomly distributed.~\cite{tiny} 
The time spectra are well fitted with Eqs. (\ref{eq1}) and (\ref{eq2}), as shown by solid lines in Figs. 1 and 2. 

\begin{figure}[tbp]
\begin{center}
\includegraphics[width=1.0\linewidth]{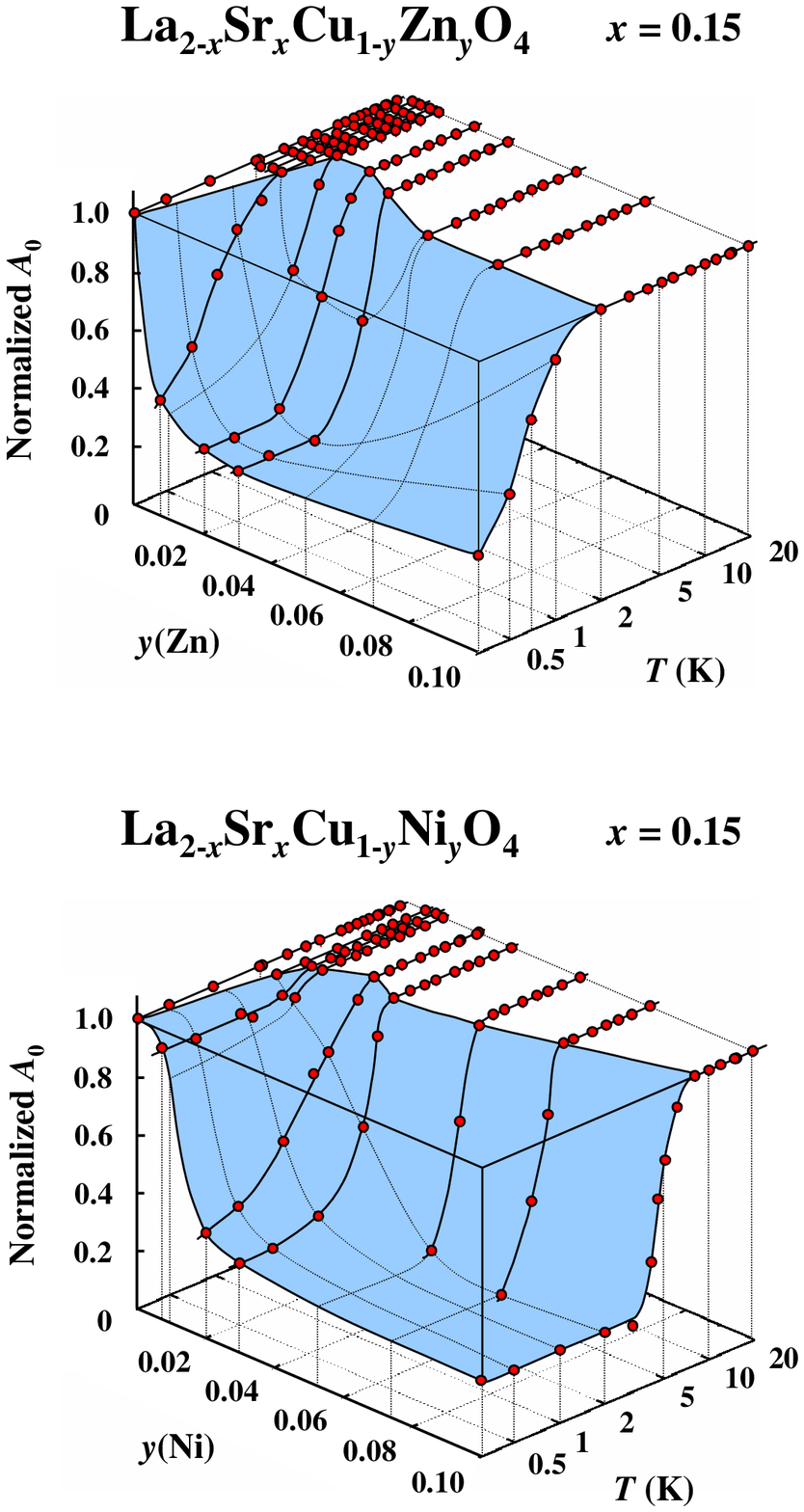}
\end{center}
\caption{(color online) Temperature dependence of the initial asymmetry of the slowly depolarizing component $A_0$, normalized by its value at a high temperature of 15 K or 20 K, for La$_{2-x}$Sr$_x$Cu$_{1-y}$(Zn,Ni)$_y$O$_4$ with $x=0.15$. The shaded area at low temperatures corresponds to the state in which the Cu-spin fluctuations slow down and/or a magnetic order is formed. Solid and dotted lines are to guide the reader's eye.}  
\label{fig:fig3} 
\end{figure}

Figure 3 shows the temperature dependence of $A_0$ normalized by its value at a high temperature of 15 K or 20 K for La$_{2-x}$Sr$_x$Cu$_{1-y}$(Zn,Ni)$_y$O$_4$ with $x=0.15$. 
The temperature dependence of $A_0$ is often used as a probe of the magnetic transition, because it reflects the volume fraction of the nonmagnetic region.~\cite{torikai,nabe1,nabe2}
In fact, $A_0 = 1$ means that the spectrum is represented only with the first term of Eq. (\ref{eq1}) or (\ref{eq2}), indicating that all the Cu spins fluctuate fast beyond the $\mu$SR frequency window. 
On the contrary, $A_0 \sim 1/3$ indicates that all the Cu spins are in the short- or long-range static magnetically ordered state. 
In Fig. 3, the shaded area at low temperatures corresponds to the state in which the Cu-spin fluctuations slow down and/or a magnetic order is formed. 
For the Zn substitution, $A_0$ decreases around $y$(Zn) $=0.01 - 0.03$ at 2 K, while the area in which $A_0$ decreases expands at low temperatures below 2 K, indicating that the Cu-spin correlation is developed with decreasing temperature in the Zn-substituted samples. 
Focusing on the temperature, $T_{\rm N}^{\rm onset}$, at which $A_0$ starts to decrease with decreasing temperature, it increases with increasing $y$(Zn) and shows the maximum around $y$(Zn) $=0.01 - 0.02$ and decreases above $y$(Zn) $>0.02$, suggesting that the Cu-spin correlation develops most around $y$(Zn) $=0.01 - 0.02$.
In fact, $A_0$ is almost 1/3 below $\sim$ 1 K for $y$(Zn) $=0.02$ and 0.03, indicating a static magnetic state. 
For $y$(Zn) $>0.03$, the area in which $A_0$ decreases shrinks with increasing $y$(Zn), but $A_0$ is below the unity at low temperatures due to the random magnetism with magnetic moments slightly larger than nuclear moments even for $y$(Zn) $=0.10$. 

For the Ni substitution, on the other hand, the decrease in $A_0$ becomes marked with increasing $y$(Ni) at 2 K and the area in which $A_0$ decreases expands at low temperatures below 2 K. 
These indicate that the Cu-spin correlation is developed at low temperatures also in the Ni-substituted samples. 
As for $T_{\rm N}^{\rm onset}$, it increases with increasing $y$(Ni) and shows a local maximum around $y$(Ni) $=0.02$ and changes to decrease above $y$(Ni) $>0.02$, which is similar to the Zn-substituted case.
For $y$(Ni) $>0.03$, however, $T_{\rm N}^{\rm onset}$ increases progressively with increasing $y$(Ni). 
In fact, $A_0$ at low temperatures below 2 K is about 1/3 for $y$(Ni) $\ge 0.07$, indicating the formation of a static magnetic state.

It is found in the lightly substituted samples of $y \le 0.03$ that the decrease in $A_0$ with decreasing temperature is smaller in the Ni substitution than in the Zn substitution, suggesting that Zn is more effective for the development of the magnetic order than Ni.
In the heavily substituted samples of $y>0.03$, the Cu-spin correlation tends to be destroyed by Zn while it tends to be developed by Ni.

\begin{figure}[tbp]
\begin{center}
\includegraphics[width=1.0\linewidth]{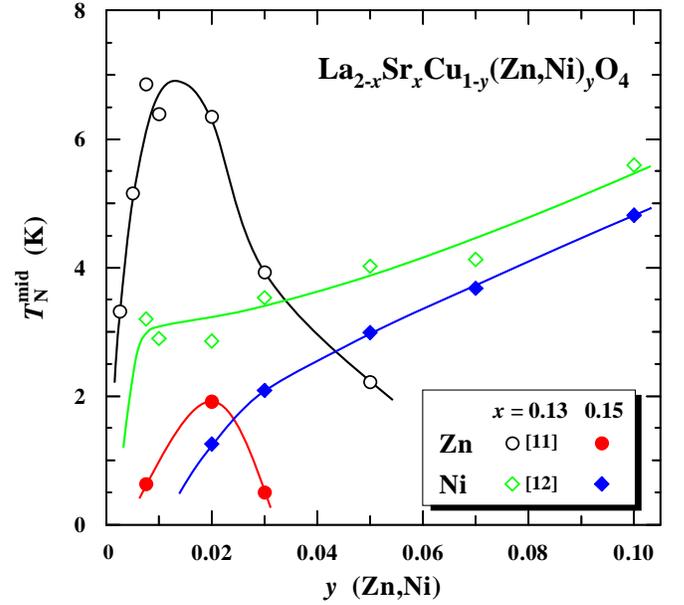}
\end{center}
\caption{(color online) Impurity-concentration dependence of $T_{\rm N}^{\rm mid}$ defined as the midpoint of the change of the $A_0$ value from the unity to the averaged value of $A_0$ in the ordered state at 2 K for La$_{2-x}$Sr$_x$Cu$_{1-y}$(Zn,Ni)$_y$O$_4$ with $x=0.15$. The data of $x=0.13$ are also plotted for comparison.~\cite{ada-prbz,ada-prbn} Solid lines are to guide the reader's eye.}  
\label{fig:fig4} 
\end{figure}

Figure 4 shows the impurity-concentration dependence of the magnetic transition temperature, $T_{\rm N}^{\rm mid}$, defined as the midpoint of the change of the $A_0$ value from the unity to the averaged value of $A_0$ in the static magnetic state at 2 K,~\cite{ada-prbz} in La$_{2-x}$Sr$_x$Cu$_{1-y}$(Zn,Ni)$_y$O$_4$ with $x=0.15$. 
The data of $x=0.13$ are also plotted for comparison.~\cite{ada-prbz,ada-prbn}
In the Zn-substituted samples with $x=0.15$, $T_{\rm N}^{\rm mid}$ appears at $y$(Zn) $=0.0075$ and shows the maximum at $y$(Zn) $=0.02$, followed by the disappearance for $y$(Zn) $>0.03$. 
In the Ni-substituted samples with $x=0.15$, on the other hand, $T_{\rm N}^{\rm mid}$ appears at $y$(Ni) $=0.02$ and increases monotonically with increasing $y$(Ni). 
The $y$ dependence of $T_{\rm N}^{\rm mid}$ in $x=0.15$ is qualitatively similar to that in $x=0.13$ in both cases of Zn and Ni substitution. 

%*****************************************************************************************
\subsubsection{Magnetic susceptibility}
%*****************************************************************************************
\begin{figure}[tbp]
\begin{center}
\includegraphics[width=1.0\linewidth]{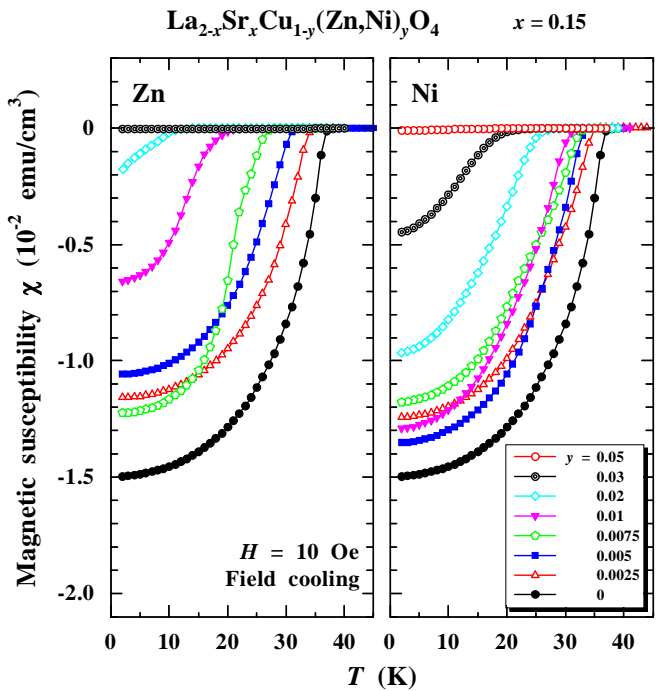}
\end{center}
\caption{(color online) Temperature dependence of the magnetic susceptibility $\chi$ for La$_{2-x}$Sr$_x$Cu$_{1-y}$(Zn,Ni)$_y$O$_4$ with $x=0.15$ and $y \le 0.05$ in a magnetic field of 10 Oe on field cooling.}  
\label{fig:fig5} 
\end{figure}

Figure 5 shows the temperature dependence of $\chi$ on field cooling for La$_{2-x}$Sr$_x$Cu$_{1-y}$(Zn,Ni)$_y$O$_4$ with $x=0.15$ and $y \le 0.05$. 
The $T_{\rm c}$ decreases with increasing $y$ in both Zn- and Ni-substituted samples and disappears at $y$(Zn) $=0.03$ and $y$(Ni) $=0.05$, respectively. 
Moreover, it is found that the absolute value of $\chi$ at 2 K, $|\chi_{\rm 2 K}|$, corresponding to the SC volume fraction decreases markedly with increasing $y$ in both Zn- and Ni-substituted samples and the decrease is more marked in the former than in the latter.
These results suggest that the SC volume fraction decreases more markedly through the Zn substitution than through the Ni substitution, which is analogous to the results in $x=0.13$.~\cite{ada-prbn}

%*****************************************************************************************
\subsubsection{Volume fractions of SC and magnetic regions}\label{volume}
%*****************************************************************************************
We estimated volume fractions of different states of Cu spins in one sample related to the three terms in Eq. (\ref{eq1}) and two terms in Eq. (\ref{eq2}), using the best-fit values of $A_0$, $A_1$, $A_2$ and $A_{\sigma}$. 
Details of the estimation of the volume fractions were described in our previous paper.~\cite{ada-prbz}
It is noted in the estimation that the region expressed by $A_1$, corresponding to the region in which the Cu-spin fluctuations slow down and/or a short-range magnetic order is formed at low temperatures, is called the $A_1$ region. 
Similarly, regions expressed by $A_0$, $A_2$ and $A_{\sigma}$ are called $A_0$, $A_2$ and $A_{\sigma}$ regions, respectively.
Volume fractions of $A_0$, $A_1$, $A_2$ and $A_{\sigma}$ regions are denoted as $V_{\rm A_0}$, $V_{\rm A_1}$, $V_{\rm A_2}$ and $V_{\rm A_{\sigma}}$, respectively. 
Figure 6 displays the impurity-concentration dependence of $V_{\rm A_0}$, $V_{\rm A_1}$, $V_{\rm A_2}$ and $V_{\rm A_{\sigma}}$ at 0.3 K for La$_{2-x}$Sr$_x$Cu$_{1-y}$(Zn,Ni)$_y$O$_4$ with $x=0.15$. 
For the Zn substitution, $V_{\rm A_0}$ rapidly decreases with increasing $y$(Zn) and becomes almost zero at $y$(Zn) $=0.03$. 
Instead, $V_{\rm A_1}$ increases and becomes almost 100 \% at $y$(Zn) $=0.03$, indicating that the Cu-spin fluctuations slow down and/or the short-range magnetic order is formed in the whole area of the sample of $y$(Zn) $=0.03$.
For $y$(Zn) $=0.10$, $V_{\rm A_0}$ is not zero but about 10 \% and almost all area of the sample is covered with the $A_{\sigma}$ region.

For the Ni substitution, on the other hand, the decrease in $V_{\rm A_0}$ is weak at $y$(Ni) $=0.0075$, while $V_{\rm A_0}$ rapidly decreases for $y$(Ni) $> 0.0075$ and reaches zero at $y$(Ni) $\sim 0.05$. 
The trend that the formation of the magnetic order requires a larger amount of Ni than that of Zn is the same as in the case of $x=0.13$.~\cite{ada-prbn}
The $A_2$ region starts to appear at $y$(Ni) $=0.03-0.10$. 
In spite of large error bars of $V_{\rm A_0}$, $V_{\rm A_1}$ and $V_{\rm A_2}$, the temperature dependence of $V_{\rm A_2}$ allows us to conclude that the $A_2$ region is formed almost entirely in the sample of $y$(Ni) $=0.10$.

Next, the SC volume fraction at 2 K, $V_{\rm SC}$, was estimated from $|\chi_{\rm 2 K}|$ shown in Fig. 5. 
The value of $V_{\rm SC}$ for each $y$ is normalized by its value at $y=0$.
The impurity-concentration dependence of $V_{\rm SC}$ is plotted in Fig. 6.
For the Zn substitution, $V_{\rm SC}$ rapidly decreases with increasing $y$(Zn), changes to slowly decrease for $y$(Zn) $ \ge 0.02$ and finally disappears for $y$(Zn) $\ge 0.03$. 
The $y$ dependence of $V_{\rm SC}$ for the Ni substitution is analogous to the $y$ dependence of $V_{\rm SC}$ for the Zn substitution, though values of $V_{\rm SC}$ for $y$(Ni) $=0.0025-0.01$ are scattered around 85 \%. 
The decrease is, however, weaker through the Ni substitution than through the Zn substitution. 
In comparison between the SC and magnetic volume fractions, the $y$ dependence of $V_{\rm SC}$ and $V_{\rm A_0}$ is roughly similar to each other in qualitative viewpoint for the Zn substitution. 
For the Ni substitution, on the other hand, the tendency of the decrease of $V_{\rm SC}$ and $V_{\rm A_0}$ seems to be similar to each other. 
Therefore, at the lowest temperature of 0.3 K, the $A_0$ region is in rough correspondence to the SC region in both Zn- and Ni-substituted samples. 
These results are similar to those obtained in $x=0.13$,~\cite{ada-prbn} in which the $A_0$ region is in rough correspondence to the SC region in both Zn- and Ni-substituted samples.

\begin{figure}[tbp]
\begin{center}
\includegraphics[width=1.0\linewidth]{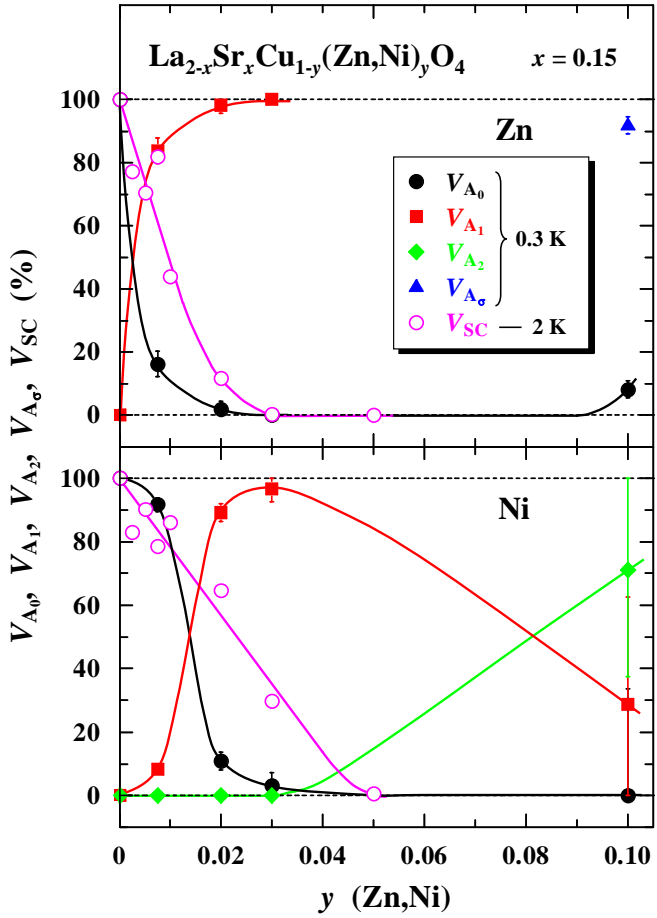}
\end{center}
\caption{(color online) Impurity-concentration dependence of $V_{\rm A_0}$ (closed circles), $V_{\rm A_1}$ (closed squares), $V_{\rm A_2}$ (closed diamonds) and $V_{\rm A_\sigma}$ (closed triangles) estimated from the $\mu$SR time spectra at 0.3 K in La$_{2-x}$Sr$_x$Cu$_{1-y}$(Zn,Ni)$_y$O$_4$ with $x=0.15$. Impurity-concentration dependence of $V_{\rm SC}$ (open circles) estimated from the magnetic susceptibility at 2 K on field cooling is also plotted. $V_{\rm A_0}$: Volume fraction of the region where Cu spins fluctuate fast beyond the $\mu$SR frequency window. $V_{\rm A_1}$: Volume fraction of the region where the Cu-spin fluctuations slow down and/or a short-range magnetic order is formed. $V_{\rm A_2}$: Volume fraction of the region where a long-range magnetic order is formed. $V_{\rm A_\sigma}$: Volume fraction of the region with slightly larger magnetic moments that nuclear dipoles being randomly distributed. $V_{\rm SC}$: Volume fraction of the superconducting region. Solid lines are to guide the reader's eye.}  
\label{fig:fig6} 
\end{figure}

%*****************************************************************************************
\subsection{Effect of Zn substitution in $x=$ 0.18 and 0.20}
%*****************************************************************************************
The ZF-$\mu$SR time spectra in La$_{2-x}$Sr$_x$Cu$_{1-y}$Zn$_y$O$_4$ with $x=0.18$ and 0.20 are shown in Fig. 1.~\cite{risdi1,risdi2} 
Focusing on the spectra at low temperatures, fast depolarization of muon spins is observed at $y$(Zn) $=0.02 - 0.03$ for $x=0.18$ and 0.20, though the depolarization is weaker than that of $x=0.15$ with $y$(Zn) $=0.02 - 0.03$. 
Such behaviors can also be confirmed by the temperature dependence of $A_0$ that $T_{\rm N}^{\rm onset}$ decreases with increasing $x$ up to $x=0.20$.~\cite{risdi2}
These indicate that the Cu-spin fluctuations slow down at low temperatures even in the overdoped regime of La$_{2-x}$Sr$_x$Cu$_{1-y}$Zn$_y$O$_4$, which is in sharp contrast with the conclusion by Panagopoulos {\it et al}. that the Zn-induced slowing down disappears above the possible quantum critical point of $x\sim 0.19$.~\cite{panago1,panago2} 
The reason for the sharp contrast is that they made $\mu$SR measurements not for $y$(Zn) $=0.03$ but for $y$(Zn) $=0.01$, 0.02 and 0.05. 

\begin{figure}[tbp]
\begin{center}
\includegraphics[width=1.0\linewidth]{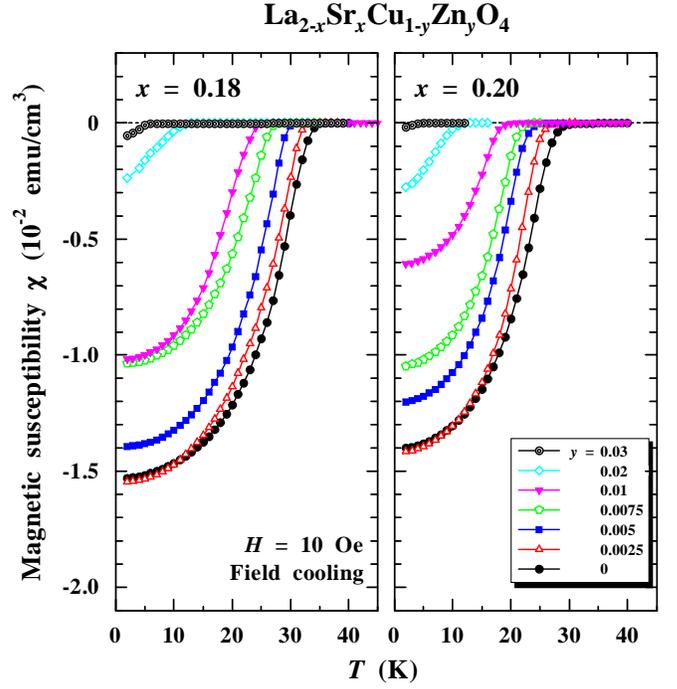}
\end{center}
\caption{(color online) Temperature dependence of the magnetic susceptibility $\chi$ for La$_{2-x}$Sr$_x$Cu$_{1-y}$Zn$_y$O$_4$ with $x=0.18$, 0.20 and $y \le 0.03$ in a magnetic field of 10 Oe on field cooling.}  
\label{fig:fig7} 
\end{figure}

Figure 7 shows the temperature dependence of $\chi$ on field cooling for La$_{2-x}$Sr$_x$Cu$_{1-y}$Zn$_y$O$_4$ with $x=0.18$, 0.20 and $y \le 0.03$. 
The $T_{\rm c}$ decreases with increasing $y$ and seems to disappear just above $y=0.03$ for both $x = 0.18$ and 0.20. 
The $|\chi_{\rm 2 K}|$, on the other hand, is found to decrease progressively with increasing $y$ for $y>0.0025$ and seems to be zero just above $y=0.03$ in both $x=0.18$ and 0.20. 
\begin{figure}[tbp]
\begin{center}
\includegraphics[width=1.0\linewidth]{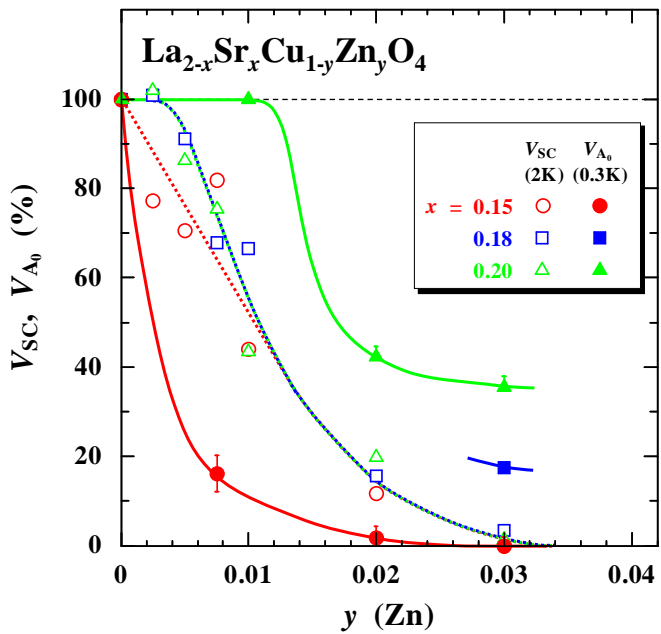}
\end{center}
\caption{(color online) Zn-concentration dependence of $V_{\rm A_0}$ (closed symbols) estimated from the $\mu$SR time spectra at 0.3 K in La$_{2-x}$Sr$_x$Cu$_{1-y}$Zn$_y$O$_4$ with $x=0.15$, 0.18 and 0.20. Zn-concentration dependence of $V_{\rm SC}$ (open symbols) estimated from the magnetic susceptibility at 2 K on field cooling is also plotted. $V_{\rm A_0}$: Volume fraction of the region where Cu spins fluctuate fast beyond the $\mu$SR frequency window. $V_{\rm SC}$: Volume fraction of the superconducting region. Solid and dotted lines are to guide the reader's eye.}  
\label{fig:fig8} 
\end{figure}
It is found in the $y$ dependence of $V_{\rm SC}$ in Fig. 8 that $V_{\rm SC}$'s of $x=0.18$ and 0.20 are roughly located on the same line. 
That is, the magnitude of the suppression of the SC volume fraction by the Zn substitution is similar to each other in the overdoped samples of $x=0.18$ and 0.20.

The Zn-concentration dependence of $V_{\rm A_0}$ at 0.3 K for $x = 0.15$, 0.18 and 0.20 is shown in Fig. 8. 
The $V_{\rm A_0}$ for $x = 0.18$ and 0.20 was estimated from fitting the spectra using Eq. (\ref{eq1}). 
With increasing $x$, the magnitude of the decrease in $V_{\rm A_0}$ with increasing $y$ becomes small.
As for the comparison between $V_{\rm SC}$ and $V_{\rm A_0}$, $V_{\rm SC}$ is smaller than $V_{\rm A_0}$ for $x=0.18$ and 0.20 and vice versa for $x=0.15$, though decreases in $V_{\rm SC}$ and $V_{\rm A_0}$ with increasing $y$ seem to be qualitatively similar to each other for $x = 0.20$ as well as for $x = 0.15$. 

%*****************************************************************************************
\section{Discussion}
%*****************************************************************************************
\subsection{Spatial distribution of SC and magnetic regions in the CuO$_2$ plane in the ground state}
%*****************************************************************************************
First, we discuss Zn- and Ni-concentration dependence of the SC and magnetic volume fractions at 0.3 K in the ground state for La$_{2-x}$Sr$_x$Cu$_{1-y}$(Zn,Ni)$_y$O$_4$ with $x=0.15$.
Figure 9 displays schematic pictures of the spatial distribution of $A_0$ and $A_1$ regions with different Cu-spin states at 0.3 K for typical $y$ values in La$_{2-x}$Sr$_x$Cu$_{1-y}$(Zn,Ni)$_y$O$_4$ with $x=0.15$, on the assumption that the magnetic order is developed around each Zn or Ni for lightly Zn- or Ni-substituted samples. 
This assumption has not yet been confirmed directly but is reasonable, referring to the results of scanning-tunneling-microscopy measurements in the Zn- or Ni-substituted Bi$_2$Sr$_2$CaCu$_{2-y}$(Zn,Ni)$_y$O$_{8+\delta}$.~\cite{pan,hudson} 
Sizes of the circles were calculated from the ratio of $V_{\rm A_0}$: $V_{\rm A_1}$ in each $y$.
It is noted from the calculation in each $y$ that the size of the circle was found to be almost independent of $y$. 
For the impurity-free sample of $y=0$, the $A_0$ region covers the whole area of the CuO$_2$ plane where the superconductivity appears. 
Through the 0.75 \% substitution of Zn for Cu, $A_1$ regions, namely, slowly fluctuating and/or short-range magnetically ordered regions of Cu spins are formed with the radius, $r_{\rm ab}$, $\sim 23$ ${\rm \AA}$ around each Zn in the $A_0$ region, namely, in the fast fluctuating sea of Cu spins. 
It is found that a large part of the CuO$_2$ plane is covered with $A_1$ regions, so that $V_{\rm SC}$ in rough correspondence to $V_{\rm A_0}$ is strongly diminished. 
In comparison between $r_{\rm ab}$ and the mean distance between Zn atoms $<$$R_{\rm Zn-Zn}$$>$ $\sim44$ ${\rm \AA}$ at $y$(Zn) $=0.0075$, $A_1$ regions overlap one another mostly. 
With increasing $y$(Zn), $A_1$ regions become dominant and the whole area of the CuO$_2$ plane is covered with $A_1$ regions for $y$(Zn) $=0.03$, indicating that the superconductivity is completely suppressed.
In the case of $x=0.13$, once $A_1$ regions start to overlap one another, $A_2$ regions corresponding to a long-range magnetically ordered regions start to be developed.~\cite{ada-prbz} 
For $x=0.15$, on the contrary, $A_2$ regions are not generated in spite of the overlap between $A_1$ regions. 
The reason may be due to a shorter-range magnetic order in $x=0.15$ than in $x=0.13$.  

For the Ni substitution, on the other hand, $A_1$ regions appear through the 0.75 \% substitution and $r_{\rm ab}$ is estimated to be $\sim 11$ ${\rm \AA}$, which is smaller than that around Zn. 
These indicate that $A_1$ regions do not overlap one another mostly. 
For $y$(Ni) $=0.03$, a large part of the CuO$_2$ plane is covered with $A_1$ regions. 
These changes with increasing $y$ for both Zn- and Ni-substituted samples suggest that the SC region competes with slowly fluctuating and/or short-range magnetically ordered regions of Cu spins.

\begin{figure}[tbp]
\begin{center}
\includegraphics[width=0.9\linewidth]{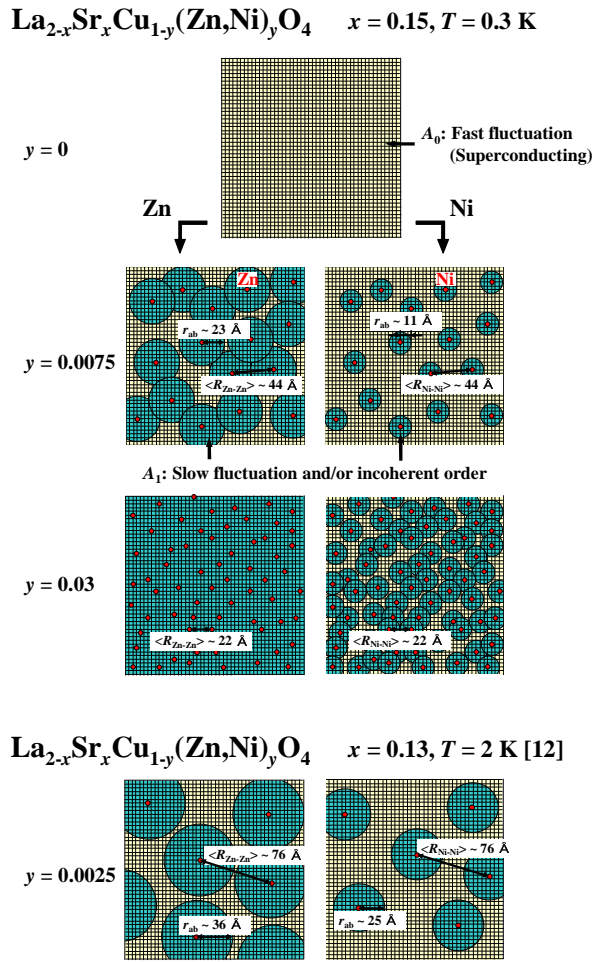}
\end{center}
\caption{(color online) Schematic pictures of the spatial distribution of different Cu-spin states in the CuO$_2$ plane, corresponding to $A_0$ and $A_1$, at 0.3 K for typical $y$ values in La$_{2-x}$Sr$_x$Cu$_{1-y}$(Zn,Ni)$_y$O$_4$ with $x=0.15$. Each crossing point of the grid pattern represents the Cu site. Zn and Ni atoms are randomly distributed in the CuO$_2$ plane. Sizes of the circles are calculated from the ratio of $V_{\rm A_0} : V_{\rm A_1}$ in each $y$. The $r_{\rm ab}$ indicates the radius of the $A_1$ region around each Zn or Ni. The $<$$R_{\rm Zn-Zn}$$>$ and $<$$R_{\rm Ni-Ni}$$>$ indicate mean distances between Zn atoms and between Ni atoms, respectively. Similar schematic pictures of $x=0.13$ are also shown for comparison.~\cite{ada-prbn}}  
\label{fig:fig9} 
\end{figure}

As seen in the case of $y=0.0075$, the size of each $A_1$ region is larger around Zn than around Ni, which is similar to the case of $x=0.13$ as shown in Fig. 9.~\cite{ada-prbn}
Itoh {\it et al}. have reported from NQR measurements in the Zn- and Ni-substituted YBa$_2$(Cu$_{1-y}$M$_y$)$_4$O$_8$ (M = Zn, Ni) that the size of the wipeout region where the Cu-spin correlation is developed is $\sim 6$ ${\rm \AA}$ around Zn, while in the case of Ni, it is limited just on the Ni site.~\cite{ito} 
The difference between Zn and Ni substitution is qualitatively similar to that in our present results. 
The quantitative difference between the NQR and our $\mu$SR results is probably due to the difference that the time scale of the Cu-spin fluctuations observed by NQR is longer than that by $\mu$SR. 
It is noted, on the other hand, that Ouazi {\it et al}. have reported contrasting results on the Zn- and Ni-substitution effects from NMR measurements in partially Zn- and Ni-substituted YBa$_2$Cu$_3$O$_7$.~\cite{ouazi} 
They have suggested that both Zn and Ni induce a staggered paramagnetic polarization with the typical extension $\xi=3$ cell units for Zn and $\xi \ge 3$ for Ni, which is inconsistent with our model and Ref. \cite{ito}, though the exact reason has not yet been clarified. 

As for the difference between results in $x=0.15$ and in $x=0.13$, as shown in the case of $x=0.15$ and $y=0.0075$ and the case of $x=0.13$ and $y=0.0025$ in Fig. 9, $r_{\rm ab}$ is smaller in $x=0.15$ even at 0.3 K than in $x=0.13$ at 2 K for both Zn and Ni substitution. 
These indicate that the development of the Cu-spin correlation around Zn and Ni is less in $x=0.15$ than in $x=0.13$. 
The reason is discussed in Sec. \ref{stripe}. 

%*****************************************************************************************
\subsection{Effect of Ni substitution}
%*****************************************************************************************
Nakano {\it et al}. have insisted from $\chi$ measurements in La$_{2-x}$Sr$_x$Cu$_{1-y}$Ni$_y$O$_4$ that Ni tends to be substituted as Ni$^{3+}$ (the spin quantum number $S=1/2$) instead of Ni$^{2+}$ for $y$(Ni) $\le 0.01$ in $x=0.15$.~\cite{nakano} 
Moreover, Matsuda {\it et al}.~\cite{matsuda} and Hiraka {\it et al}.~\cite{hiraka} have insisted from neutron-scattering experiments in La$_{2-x}$Sr$_x$Cu$_{1-y}$Ni$_y$O$_4$ with $x=0.05 - 0.07$ that Ni tends to be substituted as Ni$^{3+}$ or the substituted Ni$^{2+}$ tends to trap a hole and form the so-called Zhang-Rice singlet state at low temperatures in which the hole is localized, leading to the decrease in the effective hole concentration, $p_{\rm eff}$, as in the case of doped La$_2$NiO$_4$.~\cite{zaanen,pellegrin}
In any case, Ni spins are effectively regarded as $S=1/2$, suggesting that the Ni substitution for Cu does not destroy the Cu-spin correlation with $S=1/2$. 
In the present results, however, the depolarization of muon spins in the $\mu$SR time spectra becomes fast progressively with increasing $y$(Ni) as shown in Figs. 1 and 2, indicating the development of the Cu-spin correlation. 

On the other hand, one may claim that the decrease in $p_{\rm eff}$ by the Ni substitution results in transfer into the underdoped regime, leading to the development of the Cu-spin correlation. 
Supposed that each Ni traps a hole, for example, $p_{\rm eff}$ of the 2 \% Ni-substituted sample with $x=0.15$ should be in correspondence to that of the Ni-free sample with $x=0.13$. 
This is, however, inconsistent with the result that the former shows a fast depolarization of muon spins as shown in Fig. 1 while the latter does not at 2 K.~\cite{ada-prbn} 
Therefore, it appears that Ni does not necessarily trap one hole exactly for the optimally doped $x=0.15$ in comparison with $x=0.05 - 0.07$.~\cite{matsuda,hiraka} 
In any case, further experiments are necessary to draw a conclusion.

%*****************************************************************************************
\subsection{Relation to the spin-charge stripe order}\label{stripe}
%*****************************************************************************************
As mentioned in Sec. \ref{intro}, in the case of $x=0.13$, it has been concluded that non-SC regions around Zn and Ni are in correspondence to regions where the dynamical stripe correlations of spins and holes are pinned and localized to be a static stripe order by Zn and Ni.~\cite{ada-prbn} 
For the Zn substitution, $r_{\rm ab}$ estimated at 2 K shows the maximum around $x=0.115$ in $x = 0.10 - 0.13$.~\cite{ada-prbz} 
As mentioned above, moreover, $r_{\rm ab}$ is smaller in $x=0.15$ even at 0.3 K than in $x=0.13$ at 2 K. 
This is also the case for the Ni substitution. 
This nonmonotonic $x$ dependence of $r_{\rm ab}$ eliminates the explanation that the magnetically ordered regions around impurities become small due to weakening of the Cu-spin correlation by doped holes. 
Rather, this result suggests that the dynamical stripe correlations are pinned and localized around Zn and Ni, leading to the formation of the static stripe order. 
This is because the dynamical stripe correlations develop most around $x=0.115$.~\cite{yamada} 
In fact, from elastic neutron-scattering measurements in La$_{2-x}$Sr$_x$Cu$_{1-y}$(Zn,Ni)$_y$O$_4$ with $x=0.15$,~\cite{kofu} incommensurate elastic magnetic peaks corresponding to the formation of the static spin stripe order have been confirmed for $y$(Zn) $ = 0.017$ and for $y$(Ni) $=0.029$. 

To be summarized, the present results can be understood by the concept of the stripe pinning that the dynamical stripe correlations are pinned and localized around Zn and Ni, so that the static stripe order is formed and frequencies of the Cu-spin fluctuations come into the $\mu$SR frequency window at low temperatures.
The difference in size of the non-SC and stripe-pinned region between Zn- and Ni-substituted samples can be interpreted as described precisely in the former paper,~\cite{ada-prbn} taking into account the difference in the total energy of the superexchange energy between Cu spins and the transfer integral of holes between two cases; one is the case that Zn or Ni is located at a charge (hole) stripe and the other is the case that Zn or Ni is at a spin stripe. 

%*****************************************************************************************
\subsection{Comparison with neutron-scattering results}
%*****************************************************************************************
Kimura {\it et al}.~\cite{kimura} and Kofu {\it et al}.~\cite{kofu} have reported from inelastic neutron-scattering experiments in La$_{2-x}$Sr$_x$Cu$_{1-y}$(Zn,Ni)$_y$O$_4$ with $x=0.15$ that a gap opens in the low-energy spin excitation spectrum for the impurity-free sample of $y=0$, while an in-gap state appears through the Zn substitution so that incommensurate elastic magnetic peaks are observed for $y$(Zn) $\ge 0.017$. 
For the Ni substitution, on the other hand, the gap energy decreases with increasing $y$(Ni) and incommensurate elastic magnetic peaks are observed for $y$(Ni) $\ge 0.029$. 
Our present results that the formation of the magnetic order requires a larger amount of Zn than that of Ni are consistent with their neutron-scattering results that elastic magnetic peaks start to be observed at $y$(Zn) $=0.017$ and $y$(Ni) $=0.029$. 
On the other hand, slowing down of the Cu-spin fluctuations is observed below 2 K in the $\mu$SR time spectra for $y$(Zn) $=0.0075$, while incommensurate elastic magnetic peaks are unobservable down to 1.5 K in the neutron-scattering measurements for $y$(Zn) $\sim0.0075$. 
These suggest that the Cu-spin correlation is locally developed only around Zn or Ni and are detectable by means of $\mu$SR being able to probe the local Cu-spin correlation. 
This is consistent with our proposed model shown in Fig. 9. 

%*****************************************************************************************
\subsection{Relation between the Cu-spin dynamics and superconductivity in the overdoped regime}
%*****************************************************************************************
For the Zn-substituted samples of $x \le 0.15$, $V_{\rm A_0}$ and $V_{\rm SC}$ are in rough correspondence to each other, suggesting that the non-SC region around Zn roughly corresponds to the region where the dynamical stripe correlations are pinned and stabilized.~\cite{ada-prbz} 
In the overdoped regime of $x=0.20$, as shown in Fig. 8, the $y$ dependence of $V_{\rm A_0}$ and $V_{\rm SC}$ seems to be similar to each other from the qualitative viewpoint. 
These suggest that the superconductivity is realized in a region in which Cu-spins fluctuate fast beyond the $\mu$SR frequency window even in the overdoped regime. 
It is noted for $x=0.20$ that $V_{\rm A_0}$ starts to decrease with increasing $y$(Zn) above $y$(Zn) $=0.01$, suggesting that more than 1 \% Zn is required for the pinning and stabilization of the dynamical stripe correlations due to weakening of the Cu-spin correlation in the overdoped regime. 

It is found that $V_{\rm SC}$ is larger than $V_{\rm A_0}$ in each $y$ for $x=0.15$. 
In the overdoped regime of $x=0.20$, on the other hand, $V_{\rm SC}$ is smaller than $V_{\rm A_0}$ in each $y$. 
One may claim that in the overdoped regime, the development of the Cu-spin correlation tends to be weakened and lower temperatures than 0.3 K are needed to observe the development of the Cu-spin correlation. 
We have performed, however, ZF-$\mu$SR measurements for $x=0.20$ and $y$(Zn) $=0.03$ down to 0.02 K and found almost no development of the Cu-spin correlation more than 0.3 K.~\cite{risdi1,risdi2} 
The smaller $V_{\rm SC}$ than $V_{\rm A_0}$ in the overdoped regime of $x=0.20$ is possibly explained by the inhomogeneity of superconductivity. 
Formerly, transverse-field $\mu$SR experiments have revealed the decrease in the SC carrier density with increasing hole concentration in the overdoped regime of Tl$_2$Ba$_2$CuO$_{6+\delta}$.~\cite{ps-uemura,ps-nieder,ps-uemura2}
Moreover, $\chi$ measurements have revealed the decrease in the SC volume fraction with increasing hole concentration in the overdoped regime of La$_{2-x}$Sr$_x$CuO$_4$.~\cite{ps-tanabe1,ps-tanabe2,ps-adachi,ps-tanabe3,ps-tanabe4} 
These suggest the occurrence of a phase separation into SC and normal-state regions in the overdoped high-$T_{\rm c}$ cuprates. 
For $x = 0.20$, it is considered that a small amount of non-SC regions are formed in the CuO$_2$ plane.
In this case, due to the reduction of the SC volume fraction in the phase-separated state, it is reasonable that $V_{\rm SC}$ is smaller than $V_{\rm A_0}$ in each $y$ for $x=0.20$. 

To be summarized, it appears in the overdoped regime where the phase separation into SC and normal-state regions takes place that the dynamical stripe correlations are pinned and localized around Zn in the SC region, resulting in the decrease in $V_{\rm A_0}$ and $V_{\rm SC}$. 
With increasing $x$, owing to the decrease in the SC volume fraction and weakening of the Cu-spin correlation, it is possible that the volume fraction of the slowly fluctuating and/or short-range magnetically ordered region of Cu spins decreases and finally reaches zero at the boundary between the SC and normal-state phases in La$_{2-x}$Sr$_x$CuO$_4$.~\cite{risdi1,risdi2,wakimoto}

%*****************************************************************************************
\section{Summary}
%*****************************************************************************************
Effects of Zn and Ni impurities on the Cu-spin dynamics and superconductivity have been investigated from the ZF-$\mu$SR and $\chi$ measurements for the optimally doped and overdoped La$_{2-x}$Sr$_x$Cu$_{1-y}$(Zn,Ni)$_y$O$_4$ with $x = 0.15$, 0.18 and 0.20, changing $y$ up to 0.10 in fine step. 
In the optimally doped $x = 0.15$, the ZF-$\mu$SR measurements have revealed that, for both Zn- and Ni-substituted samples, the Cu-spin fluctuations start to exhibit slowing down at $y = 0.0075$ with increasing $y$. 
The depolarization of muon spins is, however, faster in the Zn-substituted sample than in the Ni-substituted one. 
These results indicate that the formation of a magnetic order requires a larger amount of Ni than that of Zn, which is similar to our previous results for $x = 0.13$. 
The $\chi$ measurements have revealed that the SC volume fraction strongly decreases by a small amount of Zn and the decrease is more marked than by a small amount of Ni, which is also similar to our previous results for $x = 0.13$. 

From the estimation of volume fractions of the SC and magnetic regions for $x=0.15$, it has been found that the size of each region where Cu-spin fluctuations exhibit slowing down is larger around Zn than around Ni, which is similar to the case of $x=0.13$. 
Moreover, it has been found that the region where Cu spins fluctuate fast beyond the $\mu$SR frequency window is in rough correspondence to the SC region in both Zn- and Ni-substituted samples. 
According to the concept of the stripe pinning, it follows that the dynamical stripe correlations of spins and holes are pinned and localized around Zn and Ni even for $x = 0.15$, leading to the formation of the static stripe order and the suppression of superconductivity. 
These may indicate an importance of the dynamical stripe in the appearance of the high-$T_{\rm c}$ superconductivity in the hole-doped cuprates. 

As for the overdoped regime of $x=0.18$ and 0.20, Zn-induced slowing down of the Cu-spin fluctuations has also been observed for $y$(Zn) $=0.02 - 0.03$, though the slowing down has been weakened progressively with an increase in $x$. 
These suggest the occurrence of the stripe pinning even in the overdoped regime. 
In comparison between volume fractions of SC and magnetic regions, the region where Cu spins fluctuate fast beyond the $\mu$SR frequency window seems to be in rough correspondence to the SC region even in the overdoped regime. 
It appears that a pinning and stabilization of the dynamical stripe correlations takes place in the SC regions in the phase-separated state into SC and normal-state regions in the overdoped high-$T_{\rm c}$ cuprates.

%*****************************************************************************************
\section*{Acknowledgments}
%*****************************************************************************************
We are grateful to Prof. K. Nagamine for the continuous encouragement. 
This work was partly supported by Joint Programs of the Japan Society for the Promotion of Science, by a TORAY Science and Technology Grant and also by a Grant-in-Aid for Scientific Research from the Ministry of Education, Culture, Sports, Science and Technology, Japan.


\begin{references}
\bibitem{tiny}
On the analogy of the results for $y$(Zn) $=0.10$ in $x=0.13$,~\cite{ada-prbz} a magnetic order is destroyed by a large amount of nonmagnetic Zn at $y$(Zn) $=0.10$ in $x=0.15$. In this case, Cu spins no longer produce large magnetic moments corresponding to an exponential-like fast depolarization of muon spins but might produce slightly larger magnetic moments than nuclear moments, leading to Gaussian-like depolarization with a rather narrow width.
\end{references}
\end{document}